          [2001/04/25 1.1 (PWD)]
\documentclass[a4paper,twocolumn]{esapub} 
\usepackage{natbib}
\usepackage{epsf, psfig, epsfig}

\title{GRS 1915$+$105: The first three months with {\it INTEGRAL}}
\author[1]{D.C. Hannikainen}
\author[1]{O. Vilhu}
\author[2,3]{J. Rodriguez}
\author[4]{N.J. Westergaard}
\author[3,5]{S. Shaw}
\author[6]{G.G. Pooley}
\author[7]{T. Belloni}
\author[8]{A.A. Zdziarski}
\author[9]{R.W. Hunstead}
\author[10]{K. Wu}
\author[4]{S. Brandt}
\author[11]{A. Castro-Tirado}
\author[5]{P.A. Charles}
\author[5]{A.J. Dean}
\author[2]{Ph. Durouchoux}
\author[12]{R.P. Fender}
\author[1]{P. Hakala} 
\author[5]{C.R. Kaiser}
\author[13]{A.R. King}
\author[4]{N. Lund}
\author[2]{I.F. Mirabel}
\author[14]{J. Poutanen}
  \affil[1]{Observatory, PO Box 14, FIN-00014 University of Helsinki,
  Finland, diana@astro.helsinki.fi}
  \affil[2]{Centre d'Etudes de Saclay, DAPNIA/Service d'Astrophysique,
  Gif-sur-Yvette Cedex 91191, France}
\affil[3]{{\it INTEGRAL} Science Data Center, Chemin d'Ecogia 16, CH-1290 Versoix, Switzerland}
  \affil[4]{Danish Space Research Institute, Juliane Maries Vej 30, Copenhagen O, DK-2100 Denmark}
  \affil[5]{Dept. of Physics and Astronomy, University of Southampton,
   Southampton SO17 1BJ, UK}
  \affil[6]{Astrophysics Group, Cavendish Laboratory, University of
  Cambridge, Cambridge CB3 0HE, UK}
  \affil[7]{INAF - Osservatorio Astronomico di Brera, via E. Bianchi 46, 23807 Merate (LC), Italy}
 \affil[8]{Nicolaus Copernicus Astronomical Center, Bartycka 18, 00-716 Warszawa, Poland} 
 \affil[9]{School of Physics, University of Sydney, NSW 2006, Australia}
  \affil[10]{MSSL, University College London, Holmbury St. Mary, Surrey, RH5 6NT, UK}
 \affil[11]{Instituto de Astrof\'{\i}sica de Andaluc\'{\i}a
  (IAA-CSIC), PO Box 03004, 18080 Granada, Spain}
  \affil[12]{Astronomical Institute ``Anton Pannekoek'', University of
  Amsterdam, Amsterdam, Netherlands}
  \affil[13]{Theoretical Astrophysics Group, University of Leicester,
  Leicester LE1 7RH, UK}   
   \affil[14]{Astronomy Division, P.O.Box 3000, FIN-90014 University of
  Oulu, Finland}

\begin{document}
\def\grs1915{GRS~1915$+$105}


\maketitle


\begin{abstract}
\grs1915 is being observed as part of an Open Time monitoring
  program with {\it INTEGRAL}.
Three out of six observations from the monitoring program 
  are presented here, in addition to data obtained 
  through an exchange with other observers.
We also present simultaneous {\it RXTE} observations of \grs1915.
During {\it INTEGRAL} Revolution 48 (2003 March 6) the source was observed to be in a
  highly variable state, characterized by 5-minute quasi-periodic
  oscillations. 
During these oscillations, the rise is faster than the decline,
  and is harder.
This particular type of variability has never been observed before. 
During  subsequent {\it INTEGRAL} revolutions (2003 March--May), the source 
   was in a steady or ``plateau'' state (also known as class 
  $\chi$ according to Belloni et al. 2000).
Here we discuss both the temporal and spectral characteristics of the
  source during the first three months of observations. 
The source was clearly detected with all three gamma-ray and X-ray
  instruments onboard {\it INTEGRAL}. 
\end{abstract}

\section{Introduction}

GRS~1915$+$105 has been extensively observed at all wavelengths ever
  since its discovery.
It was originally detected as a hard X-ray source with
  the WATCH all-sky monitor on the GRANAT satellite (Castro-Tirado,
  Brandt  \& Lund 1992) with
  a flux of 0.35~Crab in the 6--15~keV range (Castro-Tirado et al. 1994).
Subsequent monitoring with BATSE on CGRO showed it to be the
  most luminous hard X-ray source in the Galaxy (Paciesas et al. 1996), with
  L$_{\rm 20-100 keV}\sim3-6\times10^{38}$~erg~s$^{-1}$
  (for a distance of 12.5~kpc).
Apparent superluminal ejections have been observed from
  GRS~1915$+$105 on at least two occasions: the first
  time in 1994 with the VLA (Mirabel \& Rodr\'\i guez 1994) 
  and the second time in 1997 with MERLIN (Fender et al. 1999).
Both times the true ejection velocity was calculated to be $>0.9c$.
Following the ejections of 1997, Fender et al. (1999) gave an
  upper limit for the distance to the source of 11.2$\pm$0.8~kpc.
Recently, Chapuis \& Corbel (2004) refined the distance to \grs1915
  to be 9.0$\pm$3.0~kpc.
High optical absorption ($\geq 33$ magnitudes)
  towards GRS~1915$+$105 prevented the identification of the non-degenerate
  companion until recently.
However, using the VLT,  Greiner et al. (2001)
  identified the mass-donating star to be of spectral type K-M III.
The mass of the black hole was deduced by Harlaftis \& Greiner (2004)
  to be $14\pm4.4\,{\rm M}_{\odot}$.
The Rossi X-ray Timing Explorer ({\it RXTE}) has observed GRS~1915$+$105
  since its launch in late 1995 and has shown the source to be highly
  variable on all timescales from milliseconds to months
  (see e.g. Belloni et al. 2000; Morgan, Remillard \& Greiner 1997).
Belloni et al. (2000) categorized the variability into twelve distinct
  classes.
The source was detected up to $\sim$~700~keV during OSSE observations
  (Zdziarski et al. 2001).

GRS~1915$+$105 is being observed extensively with the
  European Space Agency's {\it
  International Gamma-Ray Astrophysical Laboratory} ({\it INTEGRAL}, Winkler
  et al. 2003) as
  part of the Core Program and also within the framework of an Open
  Time monitoring campaign.
We also have a simultaneous observing campaign with {\it RXTE} to
  concentrate on timing analysis -- these data are presented in
  Rodriguez et al. (2004a, b).
First results from Revolution 48 were described in detail elsewhere
  (Hannikainen et al. 2003).
During Revolution 48, the source was found to exhibit a new type of
  variability. 
Here we summarize the results from Revolution 48 and 
  we present the results from two other 
  observations which took place between 2003 April and May.
In addition to our dedicated observations, \grs1915 was in the 
  field-of-view of {\it INTEGRAL} observations of Aql X-1 (Molkov et
  al. 2003) and SS 433 (Cherepaschuk et al. 2003). 
We exchanged data with both the PIs of the latter observations.

\begin{table}
  \begin{center}
    \caption{Log of the {\it INTEGRAL} observations of \grs1915. 
    Data from Revolutions 49, 51, 53, 56, 58, and 60 were
    obtained from Molkov et al. (2003). Data from
    Revolution 70 obtained from Cherepaschuk et 
    al. (2003).}\vspace{1em}
    \renewcommand{\arraystretch}{1.2}
    \begin{tabular}[h]{ccc}
      \hline
      Revolution \# & Date   & Exposure time \\
                   & (2003) &  (ksec)      \\ 
      \hline
       48 & March 6  & 100 \\
       49 & March 10 &  55 \\
       51 & March 17 &  55 \\
       53 & March 23 &  55 \\
       56 & March 30 &  55 \\
       58 & April 6  &  55 \\
       59 & April 9  & 100 \\
       60 & April 12 &  55 \\
       69 & May 9    & 100 \\
       70 & May 12   & 100 \\
      \hline \\
      \end{tabular}
    \label{tab-table1}
  \end{center}
\end{table}

\section{Observations}

\subsection{{\it INTEGRAL} and {\it RXTE}}

\begin{figure}
\includegraphics[width=1.0\linewidth]{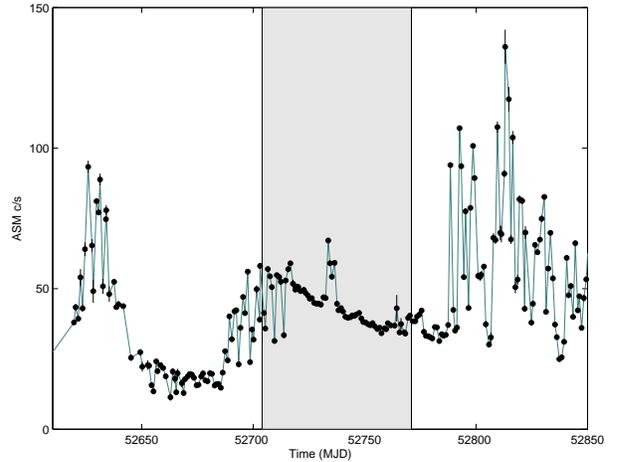}
\caption{The {\it RXTE}/ASM 1-day average 2--10~keV lightcurve. The shaded area
    indicates the dates from 2003~Mar~6 (Rev48) to May~12 (Rev70).
\label{fig-asm}}
\end{figure}

As part of a monitoring program which consists of six 
  100~ks observations separated by approximately one month, 
  {\it INTEGRAL} observed GRS~1915$+$105 for the first time for this
  campaign on 2003 Mar 6. 
Our campaign also included simultaneous coverage with the Rossi X-ray 
  Timing Experiment ({\it RXTE}) --  these data are discussed in
  Rodriguez et al. (2004a, 2004b).
Two other observations were conducted in April and May 2003, 
  and a further two in November (these latter will be discussed
  in a forthcoming paper).
Table~\ref{tab-table1} shows the log of the Open Time observations of
  \grs1915 -- these include three from the monitoring program plus
  several obtained from the data exchange with other observers.
Figure~\ref{fig-asm} shows the {\it RXTE}/ASM 2--10~keV lightcurve, and the
  shaded area indicates the dates from 2003~March~6 to May~12.


The {\it INTEGRAL} observations were undertaken using the hexagonal dither 
  pattern (Courvoisier et al. 2003) for Revolutions 48, 59 and 69:
   this consists of a hexagonal pattern around the nominal target location 
   (1 source on-axis pointing, 6 off-source pointings, each 2 degrees apart
   and each science window of 2200 s duration).
This means that GRS~1915$+$105 was always in the field-of-view
   of all three X-ray (JEM~X-2) and gamma-ray (IBIS and SPI) instruments
   throughout the whole observation.
For the other revolutions, a $5\times5$ dither pattern was used
  (Courvoisier et al. 2003); however, with the latter observing
  pattern \grs1915 is not always in the field of view of JEM~X-2.
Hence we note that 
  to fully exploit the JEM~X instrument, the hexagonal
  dither pattern is the preferred observing mode.

\subsection{Ryle Telescope}

The Ryle Telescope is an 8-element interferometer operating at 
  15~GHz (2cm wavelength). 
The elements are equatorially mounted 13 m Cassegrain antennas, on an
  (almost) E-W baseline. 
The Ryle monitors \grs1915 on a daily basis, performing several
   observations per day. 

\begin{figure}
\includegraphics[width=1.0\linewidth]{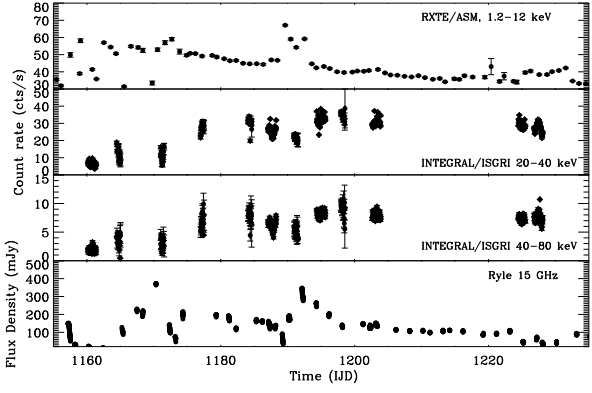}
\caption{The lightcurves from Revs. 48--70. {\it Top:} The {\it RXTE}/ASM
  lightcurve. {\it Middle:} The ISGRI 20--40 and 40--80~keV
  lightcurves. {\it Bottom:} The Ryle 15 GHz lightcurve. 
\label{fig-multi}}
\end{figure}

\subsection{Multiwavelength coverage}

Fig.~\ref{fig-multi} shows the multiwavelength lightcurves of
  GRS~1915$+$105 for the period covering 2003~Mar~6 to May~12, ie
  encompassing Revolutions 48 through 70.
The revolutions shown are tabulated in Table~\ref{tab-table1}.
In addition, we show two revolutions -- 57 and 62 -- that are dealt
  with elsewhere (see Fuchs et al. these proceedings and Fuchs et al. 2003).

\section{Revolution 48}

Although the main results from Rev 48 are presented in Hannikainen
  et al. (2003) we will summarize some of the most important points here. 

\begin{figure}
\includegraphics[width=1.0\linewidth]{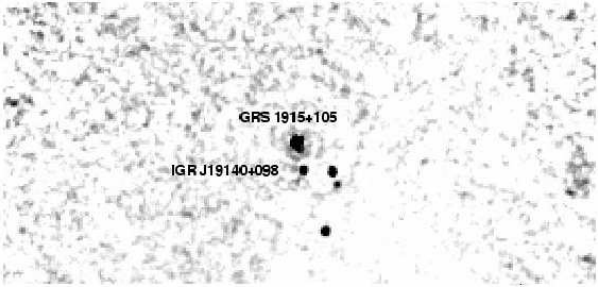}
\caption{The IBIS/ISGRI 20--40~keV ($\sim22^{\circ}$ width 
  and $\sim10^{\circ}$ 
  height) image, showing the location of GRS~1915$+$105 and three
  other bright sources in the field, including the new transient
  IGR~J1914$+$098 discovered during the observation of 2003~Mar~6. 
  North is up and East is to the left.
\label{fig-isgriimage}}
\end{figure}

Fig.~\ref{fig-isgriimage}
   shows the IBIS/ISGRI 20--40~keV
  ($\sim22^{\circ}\times\sim10^{\circ}$) mosaicked image of the field
  of \grs1915, with an exposure time of 98300~s. 
The elongation of the source is due to its apparent brightness
  in the mosaic. 
A preliminary background correction was performed (Terrier et
  al. 2003). 

A new transient was discovered during the Rev.~48 observation,
  IGR~J19140$+$098 (SIMBAD corrected name IGRJ~19140$+$0951;
  Hannikainen, Rodriguez \& Pottschmidt 2003). 
See Schultz et al. (these proceedings), Cabanac et al. (2004) 
  and Hannikainen et al. (in preparation) for details on this source.

\begin{figure}
\includegraphics[width=1.0\linewidth]{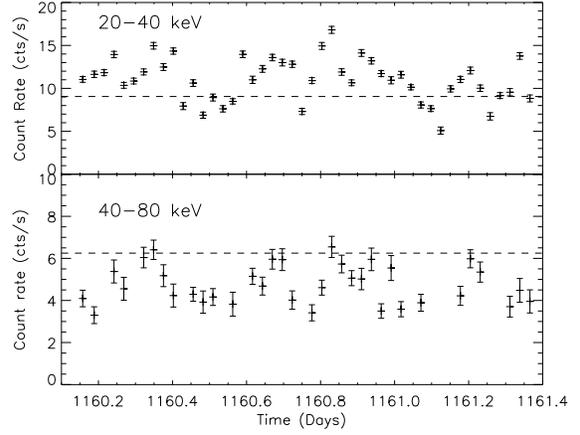}
\caption{The ISGRI lightcurves from Rev. 48. The 20--40~keV lightcurve
  (top) and the 40--80~keV lightcurve (bottom). The dashed line shows
  the 50~mCrab level in both energy ranges. The bin size is one
  science window, or 2200~s.
\label{fig-isgri}}
\end{figure}

Fig.~\ref{fig-isgri} shows that \grs1915 was highly variable during
  Rev. 48. 
The dashed horizontal line in both panels indicates the 50~mCrab level
  in the given energy ranges. 
As can be inferred, the luminosity of the source varies between
  $\sim30-100$~mCrab in the 20--40~keV range, and between $\sim24-50$~
  mCrab in the 40--80~keV range.

\begin{figure}[h]
\begin{tabular}{c}
\includegraphics[width=1.0\linewidth]{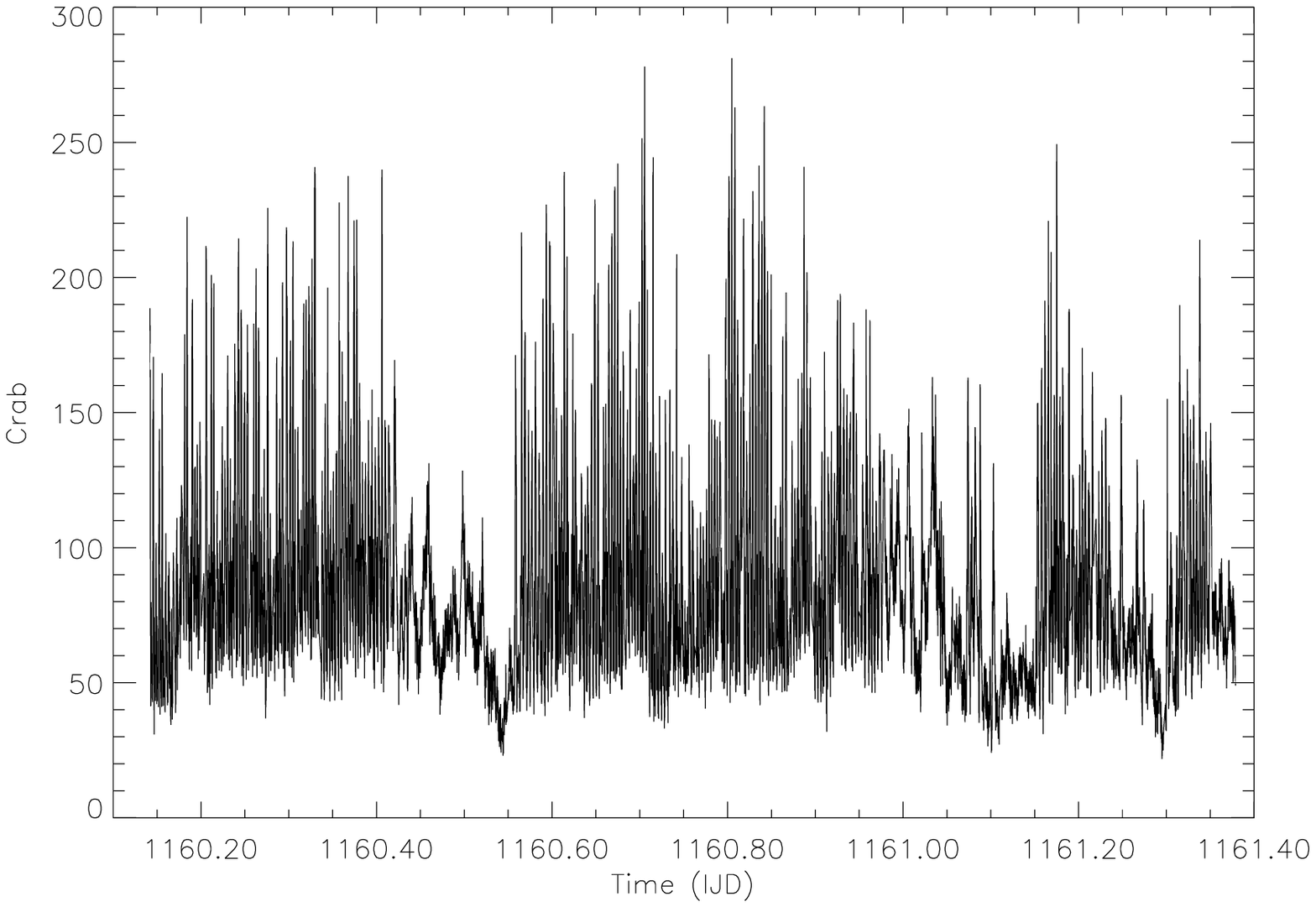} \\
\includegraphics[width=1.0\linewidth]{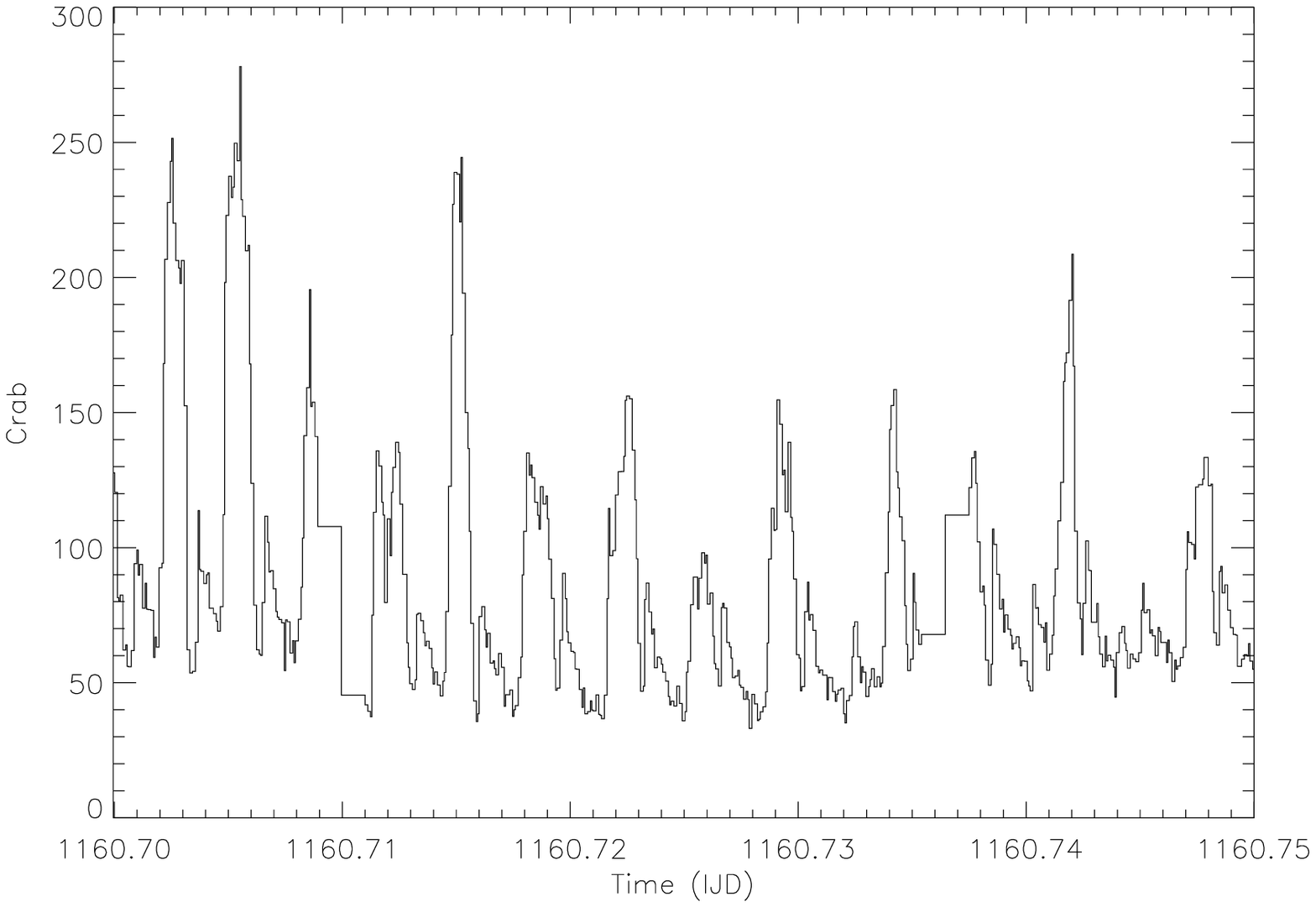} \\
\end{tabular}
\caption{{\it Top.} The JEM~X-2 lightcurve from Rev. 48, showing how
  the source varied between 0.15 and 2 Crab.
{\it Bottom.} A zoom of the JEM~X-2 lightcurve showing the 5-min
  QPOs. The bin size is 8~s.
\label{fig-jemx}}
\end{figure}

\begin{figure}
\includegraphics[width=7cm]{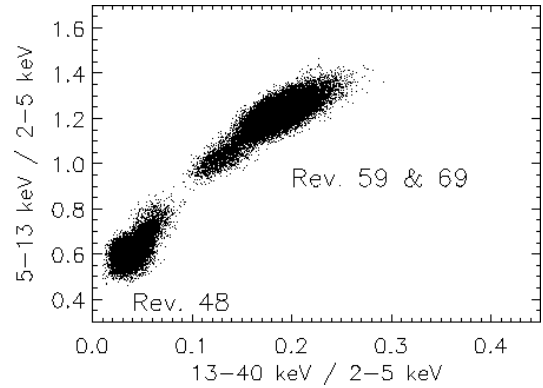}
\caption{The {\it RXTE}/PCA color-color diagram for Revs. 48, 59 and 69. 
   The time bin is 1s. The x- and y-scales correspond to those used for
    the $\chi$ classes in Fig. 2C in Belloni et al. (2000).  
\label{fig-color}}
\end{figure}

The Joint European X-ray monitor, JEM-X (Lund et al. 2003) consists of
  two identical coded mask instruments -- however, during the time of
  these observations, only JEM~X-2 was operational.
Fig.~\ref{fig-jemx} shows the JEM~X-2 lightcurve from the whole of
  Rev.~48 (top), and a zoom covering 1.2 hours (bottom). 
As can be seen, the source was in a very variable state, with the flux varying 
  between 0.25 to 2 Crab with a mean of $\sim0.5$~Crab.
The zoomed plot shows   particularly striking rapid oscillations.
Of particular interest are the main peaks separated by $\sim 5$
minutes. Although this kind of variability  resembles  the 
$\rho$-heartbeat, $\nu$ and $\phi$ oscillations (variability classes
  of Belloni et al. (2000)), these are more uniform
 and occur on shorter timescales. 
Recently, Tagger et al. (2004) have shown from a work of 
   Fitzgibbon et al. (priv. comm.) that GRS~1915$+$105 seems
  to follow the same pattern of transition through all the 12
  classes. 
This may suggest that GRS~1915$+$105 simply displays a continuum of all
  possible configurations and that we caught it in an intermediate
  class between two previously known ones. 
Our {\it RXTE} observations did not cover the entire 100 ks 
  {\it INTEGRAL} data but  were simultaneous. 
They confirm the variability seen by JEM~X-2.
We  produced a color-color (CC) diagram for the {\it RXTE} data 
in the same manner as in Belloni et al. (2000) (note that the energy-channel 
conversion for PCA corresponds to epoch 5). The resultant plot is 
  shown in Fig.~\ref{fig-color}.
This CC-diagram and the JEM~X-2 lightcurve seem to indicate a new
  type of variability not seen in the 12 classes by Belloni et al. (2000).

\begin{figure}
\includegraphics[width=8cm,angle=0]{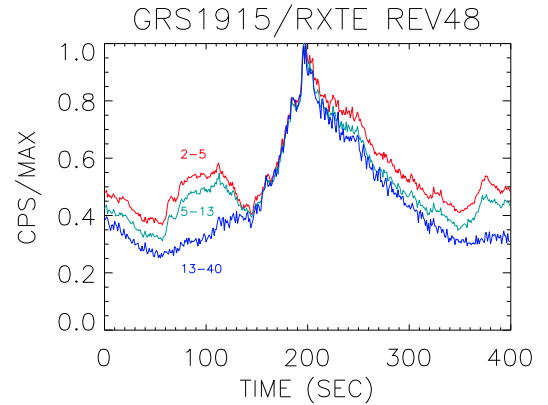}
\caption{The mean of nine {\it RXTE}/PCA pulses (2--5~keV, 
5--13~keV and 13--40~keV). The `ups' are harder than the 
`downs'. The rise is faster than the decline, which is very
unusual behavior for \grs1915.
\label{fig-mean}}
\end{figure}

Fig.~\ref{fig-mean} shows the mean of nine {\it RXTE}/PCA pulses. 
As can be seen, the `ups' were harder than the `downs'
  -- this behavior is opposite to that seen in the $\rho$, 
  $\nu$ and $\phi$ oscillations.
The rise is also faster than the decline. 
This type of behavior is very unusual for \grs1915.
In each case, there is a pre-flare not seen in the hard X-rays.

\begin{figure}
\includegraphics[width=1.0\linewidth]{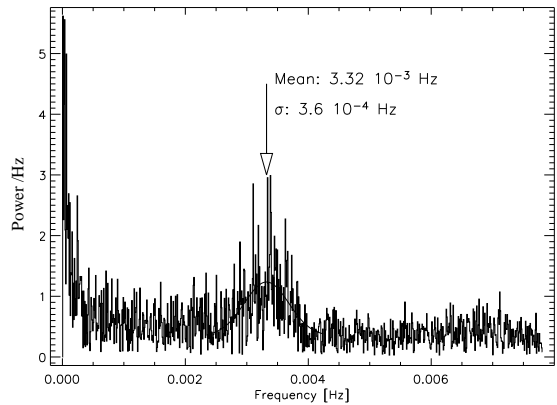}
\caption{The JEM~X-2 power spectrum. The QPO is at a frequency 
         of 3.32$\times10^{-3}$~Hz.
\label{fig-power}}
\end{figure}

In order to have a first overview of the JEM~X-2 capability in timing
   analysis, we carried out a Fourier Transform of the whole JEM~X-2 
   lightcurves and obtained a power spectrum (Fig.~\ref{fig-power}).
After removing artefacts due to some gaps in the data, we can identify
   a Quasi-Periodic Oscillation (QPO) at a frequency 
   $3.32\times10^{-3}$ Hz ($\sigma=3.6\times10^{-4}$ Hz). 
At the current stage, the exact power of the feature cannot be known
   exactly since it requires a precise background estimate.
The observation of such a low QPO with JEM~X-2 is, however, of prime
   importance since it opens the studies of low
   frequencies (difficult to access with {\it RXTE} due to its 90-minute
   orbit, and usually shorter observations) through the long,
   continuous {\it INTEGRAL} observations. 
The frequency of this  QPO is in agreement with the 5-minute timescale of the main peaks in the 
   lightcurve. 
The fact that such a feature is detected from the Fourier transform of
   the whole 100 ks JEM~X-2 lightcurve indicates that this class of variability
   is dominated by the  5-min  oscillations.
The power density spectrum was constructed for the whole 100~ksec of
   the observation and at times the oscillation disappears -- hence
   the relative amplitude of the QPO is lower. 
A deeper analysis of the variability is in progress and will be
   published elsewhere.
We conclude here that we have discovered a new type of variability in
   the JEM~X-2 lightcurve of GRS~1915$+$105.

More on the timing analysis of \grs1915 can be found in Rodriguez et
   al. (2004a, 2004b).

\section{Revolutions 49 through 70}

\begin{figure}
\includegraphics[width=1.0\linewidth]{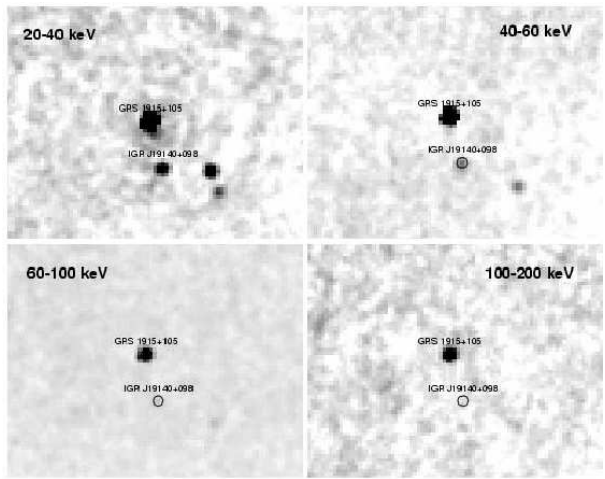}
\caption{A mosaic from Revolutions 49, 51, 53, 56, 58, 59, 60, 69 and
  70 in the 20--40~keV (top left), 40--60~keV (top right),
  60--100~keV (bottom left)  and 100--200~keV ranges (bottom
  right). The size of each image is approximately 6.15 (width)
  $\times$ 4.65 (height) degrees.
\label{fig-mosaic}}
\end{figure}

\begin{figure}
\includegraphics[width=1.0\linewidth]{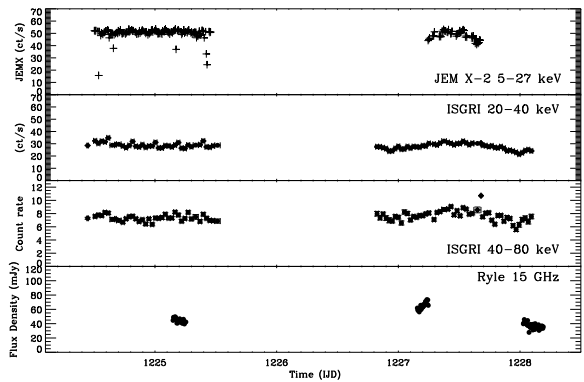}
\caption{The JEM X-2 (top), ISGRI 20--40 and 40--80~keV (middle),
and Ryle 15 GHz (bottom) lightcurves for Revs. 69 and 70. 
The JEM X-2 bin size is 500~s, while for ISGRI it is again
one science window, ie 2200~s.
\label{fig-lc69_70}}
\end{figure}

\begin{figure}
\includegraphics[width=0.75\linewidth,angle=270]{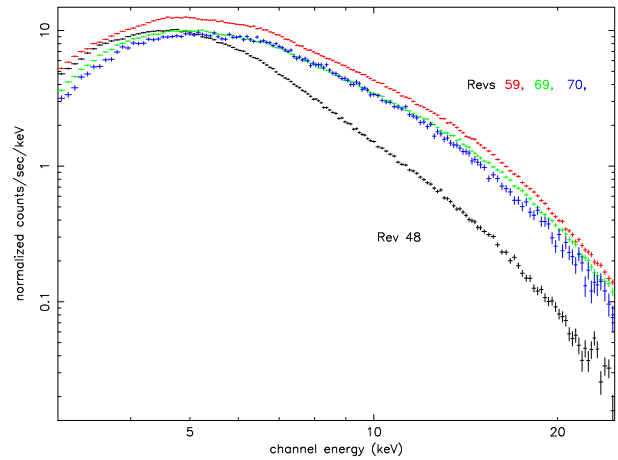}
\caption{The JEM~X-2 spectra from Revolutions 48, 59, 69 and 70.
         The figure shows how the spectrum from Rev. 48 is softer
         than those from Revs. 59, 69 and 70. However, one must 
         bear in mind that the spectrum from Rev. 48 is in fact
         an average over the extreme variability observed during
         that revolution. For clarity, the systematic errors 
         have been left out for this figure.
\label{fig-jemx_48-70}}
\end{figure}

Fig.~\ref{fig-mosaic} 
  shows the IBIS/ISGRI mosaicked image from revolutions 49,
  51, 53, 56, 58, 59, 60, 69 and 70, in the 20--40~keV, 40--60~keV,
  60--100~keV and 100-200~keV energy ranges. 
The figure shows that GRS~1915$+$105 is visible  up to 200 keV.
It also shows that the new source, IGR~J19140$+$098, is clearly detected up
   the 60--100~keV range with a significance of 6.33$\sigma$.

Fig.~\ref{fig-lc69_70} shows the lightcurves from revolutions 69 and
  70 (Rev. 59 was omitted for clarity).
The plot clearly demonstrates that the source was in a steady, ``plateau''
  state, with a constant level of emission in the ISGRI and JEM~X-2
  energy ranges. 
One can also see that the radio emission is at the $\sim
  50$~mJy level.
This is characteristic of the $\chi$ state.
More specifically, the level of radio emission allows us to identify
  the source as being in the $\chi1 - \chi3$ state, ie a {\it radio loud
  hard state} (Muno et al. 2001), also known as the {\it type II state}
  (Trudolyubov 2001).
For a comprehensive account of multiwavelength observations of the
  source in this state, see Fuchs et al. (2003) and Fuchs et al.
  (these proceedings).

Figure~\ref{fig-color} shows the color-color diagram for Rev. 59 and
  69.
It is immediately evident that Revs. 59 and 69 are virtually identical
  to each other, and that they differ significantly from Rev. 48.


\section{Spectral analysis}

\subsection{JEM~X-2: Revs 48--70}

Fig.~\ref{fig-jemx_48-70} shows the JEM~X-2 spectra from 
  Revolutions 48, 59, 69 and 70.
The spectrum from Revolution 48 is an average over the variability
  observed during that revolution. 
Data below 3~keV and above 25~keV were ignored.
The spectra were fitted with an (absorbed) disk blackbody $+$
  powerlaw $+$ Gaussian. 
Systematic errors of 20\% were added up to energy 4~keV,
  10\% to the energy range 4--7~keV, and 2\% in the range 7--25~keV.
The results of the fits are shown in Table~\ref{tab-fits}.
The column density was fixed to $N_H=5\times10^{22}\,{\rm cm}^{-2}$.
A Gaussian of centroid 6.5~keV was needed for the fits in all four
  cases.
The results show that during Rev.~48, the inner disk temperature was cooler
  than for the other revolutions.
The luminosity of the source ranges from $8.0\times10^{37}\,{\rm
  erg}\,{\rm s}^{-1}$ in Rev. 48, to $1.4\times10^{38}\,{\rm
  erg}\,{\rm s}^{-1}$ in Rev. 59, based on a distance of 9~kpc
  (Chapuis \& Corbel 2004).

\begin{table}
  \begin{center}
    \caption{Results of the spectral fits to the JEM X-2 data.}\vspace{1em}
    \renewcommand{\arraystretch}{1.2}
    \begin{tabular}[h]{ccccc}
      \hline
         Rev. \#  & $T_{in}$ & $\Gamma$ & $\chi^2_{\nu}$ & Flux$^{\dagger}$ \\
                  &  (keV)   &          &              & (3--20 keV) \\ \hline
          48 &  $2.16^{+0.26}_{-0.20}$ & $3.36^{+0.08}_{-0.08}$ & 0.93
    & 0.822 \\ 
          59 &  $3.61^{+0.22}_{-0.20}$ & $2.54^{+0.10}_{-0.09}$ & 0.70
    & 1.461\\
          69 &  $3.59^{+0.26}_{-0.21}$ & $2.45^{+0.10}_{-0.09}$ & 0.66
    & 1.178 \\
          70 &  $3.03^{+0.33}_{-0.21}$ & $2.23^{+0.27}_{-0.35}$ & 0.82
    & 1.117 \\
      \hline \\
      \end{tabular}
    \label{tab-fits}
  \end{center}
  \vspace{-8mm}
  {\footnotesize $^{\dagger}$ $\times10^{-8}$ erg cm$^{-2}$ s$^{-1}$} 
\end{table}


\subsection{JEM~X-2, ISGRI and SPI: Revs. 59--70}

\begin{figure}
\includegraphics[width=0.75\linewidth,angle=270]{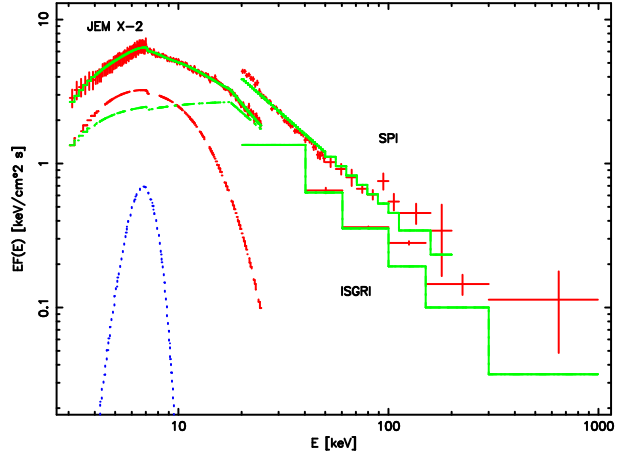}
\caption{The {\it INTEGRAL} broadband spectrum:
         JEM~X-2 + ISGRI + SPI.   
\label{fig-broad}}
\end{figure}

During Revs. 59 through 70, \grs1915 was in a steady ``plateau''
  state, as shown in Fig.~\ref{fig-lc69_70} for Revs. 69 and 70.
Figure~\ref{fig-broad} shows the resulting broadband spectrum obtained
  from JEM~X-2, ISGRI and SPI.
The discrepancy in the cross-calibration between SPI and the other two 
  instruments, JEM~X-2 and ISGRI, is a well-known fact and is taken
  into account by introducing a constant which is allowed to vary
  freely in the fitting procedure. 
The JEM~X-2 spectrum is from Rev.~59, while the ISGRI spectrum is
  extracted from the mosaic image and the SPI spectrum is co-added
  from Revs.~59, 69 and 70.
Energy ranges $<3$~keV and $>25$~keV were ignored in the JEM~X-2 data,
  $<20$~keV in the ISGRI data and $>200$~keV in the SPI data.

Two models were fitted to the data, a simple broken powerlaw, and
  an (absorbed) disk blackbody + broken powerlaw + gaussian.
Both fits yielded values comparable to those found by Trudolyubov
  (2001) for the Type II state.
However, for the broken powerlaw only model, both 
  the $\alpha_1$  and the break energy were slightly lower in our data,
  while in the disk blackbody + broken powerlaw the $\alpha_1$
  was again slightly lower but the break energy was slightly higher. 
For the disk blackbody + broken powerlaw model a gaussian was needed, 
   which was frozen to 6.5~keV.
In both cases the column density was again fixed to  
  $N_H=5\times10^{22}\,{\rm cm}^{-2}$.
In addition to the JEM~X-2 systematics, overall systematics of 5\%
  were added to the spectra.
Table~\ref{tab-models} shows the results of the fitting. 
Here we consider the simple broken powerlaw models for a direct
  comparison with Trudolyubov's (2001) results.
More physical models will be treated in future work.
For now, it is tempting to consider that the origin of the second powerlaw may
  arise in the compact jet which is detected in the radio
  (for a discussion on the compact jet see Fuchs et al. 2003
   and Rodriguez et al. 2004b). 
Fig.~\ref{fig-multi} shows that there is radio emission present
  throughout our observations at a fairly strong level, 
  which most likely arises from the compact jet. 

\begin{table}[h]
  \begin{center}
    \caption{Results of the spectral fits to the JEM X-2, ISGRI and SPI data.}\vspace{1em}
    \renewcommand{\arraystretch}{1.2}
    \begin{tabular}[h]{lc}
      \hline
Model & Broken powerlaw \\ \hline
$\alpha_1$ &  $2.07^{+0.06}_{-0.07}$ \\
Break E (keV) & $9.98^{+0.43}_{-0.21}$ \\
$\alpha_2$ & $3.23^{+0.02}_{-0.03}$ \\ 
$\chi^2_{\nu}$ & 1.40 \\ \hline
          &                        \\
Model & Disk blackbody + broken powerlaw \\ \hline
$T_{in}$ (keV) & $2.56^{+0.98}_{-0.16}$  \\
$\alpha_1$ & $1.99^{+0.50}_{-0.20}$  \\
Break E (keV) & $17.63^{+3.28}_{-0.22}$ \\ 
$\alpha_2$ & $3.30^{+0.02}_{-0.05}$ \\ 
$\chi^2_{\nu}$ & 0.85 \\ \hline
      \end{tabular}
    \label{tab-models}
  \end{center}
\end{table}

\section{Summary}
We have presented {\it INTEGRAL} results from the first three months of
  monitoring of \grs1915.
We have shown that \grs1915 exhibited two very distinct states during
  our observations: in Rev.~48 a new type of variability was
  discovered, while in Revs.~59, 69 and 70 the source was in the 
  radio loud hard state (Muno et al. 2001), also 
  known as the type II state (Trudolyubov 2001).
Spectral fits to the JEM~X-2, ISGRI and SPI data yielded parameters
  similar to those expected for the type II state, while spectral fits
  to the JEM~X-2 data alone showed the source had a high inner
  disk temperature during the ``plateau'' states. 

\section*{Acknowledgments}
DCH is a Fellow of the Academy of Finland.
JR acknowledges financial support from the French Space Agency (CNES).
This research was partly funded by the HESA/ANTARES programme
 of the Academy of Finland and TEKES, the Finnish Techonology Agency.
The authors wish to thank Dr. Sergei Molkov and Prof. Anatoly Cherepaschuk for
  generously agreeing to exchange data.
We acknowledge quick-look results from the {\it RXTE}/ASM team. 
Based on observations with {\it INTEGRAL}, an ESA project with instruments
  and science data centre funded by ESA member states (especially the PI
  countries: Denmark, France, Germany, Italy, Switzerland, Spain), Czech
  Republic and Poland, 
  and with the participation of Russia and the USA.
This research has made use of data obtained from the High Energy
  Astrophysics Science Archive Research Center (HEASARC), provided by
  NASA's Goddard Space Flight Center, and the SIMBAD database,
  operated at CDS, Strabourg, France.


\begin{thebibliography}{}
    \bibitem[Belloni et al. (2000)]{belloni}
      Belloni, T., Klein-Wolt, M., M\'endez, M., van der Klis, M.
       \& van Paradijs, J. 2000, A\&A, 355, 271

    \bibitem[Cabanac et al. (2004)]{cabanac}
     Cabanac, C., Rodriguez, J., Petrucci, P.-O., Henri, G., Hannikainen, D. \& Durouchoux, P.
     2004, astro-ph/0401308

   \bibitem[Castro-Tirado et al.(1992)]{castro}
       Castro-Tirado, A.J., Brandt, S., \& Lund, N.
                          1992, IAUC 5590

\bibitem[Chapuis \& Corbel (2004)]{chapuis}
       Chapuis, C. \& Corbel, S. 2004, A\&A, 414, 659

\bibitem[Cherepaschuk et al. 2003]{cherepaschuk}
       Cherepaschuk, A.M., Sunyaev, R.A., Sefina, E.V., Panchenko,
       I.V., Molkov, S.V. \& Postnov, K.A. 2003, A\&A, 411, L441

\bibitem[Courvoisier et al. (2003)]{courvoisier} 
     Courvoisier, T. J. L., Walter, R., Beckmann, V., et al. 2003, A\&A, 411, L53

  \bibitem[Fender et al. (1999)]{fender99} Fender, R.P., Garrington,
   S.T., McKay, D.J., et al., 1999, MNRAS, 304, 865

  \bibitem[Fuchs et al. (2003)]{fuchs}
     Fuchs, Y., Rodriguez, J., Mirabel, I.F., et al. 2003, A\&A, 409, L35

   \bibitem[Greiner et al. 2001]{greiner}
        Greiner, J., Cuby, J.G., McCaughrean, M.J., Castro-Tirado, A.
        \& Mennickent, R.E. 2001, A\&A, 373, L37

\bibitem[Hannikainen et al. (2003)]{hannikainen}
      Hannikainen, D.C., Vilhu, O., Rodriguez,J., et al. 2003, A\&A,
      411, L415 

\bibitem[Hannikainen, Rodriguez \& Pottschmidt (2003)]{hannikainenb}
         Hannikainen, D.C., Rodriguez, J.\& Pottschmidt, K. 2003, IAUC~8088


\bibitem[Harlaftis \& Greiner (2004)]{harlaftis}
        Harlaftis, E. \& Greiner, J. 2004, A\&A, 414, L13

\bibitem[Lund et al. 2003]{lund}
        Lund N., Budtz-J{\o}rgensen C., Westergaard N.J., et al, 2003,
        A\&A, 411, L231
 
\bibitem[Mirabel \& Rodr\'\i guez(1994)]{mirabel94} Mirabel, I.F. \&
       Rodr\'\i guez, L.F. 1994, Nature, 371, 46


\bibitem[Molkov et al. 2003]{molkov}
        Molkov, S., Lutovinov, A. \& Grebenev, S. 2003, A\&A, 411, L357

\bibitem[Morgan et al. 1997]{morgan}
          Morgan, E.H, Remillard, R.A. \& Greiner, J. 1997, ApJ, 482, 993

\bibitem[Muno et al. 2001]{muno} Muno, M., Remillard, R., Morgan, E., et
al., 2001, ApJ, 556, 515

\bibitem[Paciesas et al. (1996)]{paciesas}
       Paciesas, W.S, Deal, K.J., Harmon, B.A., et al. 1996, A\&AS,
       120, 205

\bibitem[Rodriguez et al. 2004]{rodriguez04} 
       Rodriguez, J., Fuchs, Y., Hannikainen, D.C., Vilhu, O., 
       Shaw, S.E., Belloni, T. \& Corbel, S. 2004a, astro-ph/0403030

\bibitem[Rodriguez et al. 2004]{rodriguez03} 
       Rodriguez, J., et al., 2004b, submitted to ApJ

\bibitem[Tagger et al. 2004]{tagger}
      Tagger, M., Varni\`ere, P., Rodriguez, J. \& Pellat, R.,
      ApJ, in press, astro-ph/0401539

\bibitem[Terrier et al. (2003)]{terrier}
       Terrier, R., Lebrun, F., Sauvageon, A., et al., 2003, A\&A, 411, L167

\bibitem[Trudolybov et al. 2001]{trudolyubov} Trudolyubov, S. 2001,
        ApJ, 558, 276

   \bibitem[Winkler et al. (2003)]{winkler}
       Winkler, C., Courvoiser, T.J.-L., DiCocco, G., et al. 2003, A\&A, 411, L1


   \bibitem[Zdziarski et al. (2001)]{zdziarski}
       Zdziarski, A.A., Grove, J.E., Poutanen, J., Rao, A.R. \&
       Vadawale, S.V. 2001, ApJ, 554, L45

\end{thebibliography}

\end{document}